# Exploring IceCube Neutrino Alerts with the HAWC Observatory

**Ian Herzog**[a,*]

[a]*Michigan State University*

*E-mail:* herzogia@msu.edu

While much work has gone into associating neutrino emission with various sources, very few sources have emerged. With the recent publication of IceCube Event Catalog (IceCat-1), the IceCube neutrino observatory has released a list of the most promising astrophysical neutrino events since operations began in 2010. Using the archival data from the High Altitude Water Cherenkov (HAWC) gamma-ray observatory, we perform a coincidence search for gamma rays and neutrinos using a Bayesian Block algorithm with the public IceCube alerts from IceCat-1 and the Astrophysical Multi-messenger Observatory Network (AMON). Of the 350 alerts considered, 25 detections were found, with 1 coinciding with the flaring HAWC source Markarian 421, an active galactic nuclei. We present the performance of this method and a discussion of physics implications.

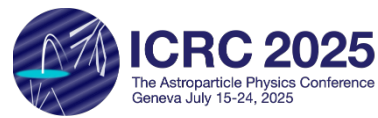

*Speaker

## 1. Introduce

Multi-messenger astronomy has proven invaluable in linking the full energy range of the electromagnetic spectrum with other particles, such as neutrinos. In particular, the TeV regime can provide constraints on the energy thresholds of these particles, as well as hinting at the mechanisms responsible [5]. With the IceCube neutrino observatory's recent detection of diffuse high energy neutrinos from the galactic plane [8], the time is ripe to look at potential coincidences of high energy neutrinos and γ-rays. IceCube will send out alerts of probable astrophysical neutrinos for other observatories to perform follow-up searches [2]. Using these alerts and the High-Altitude Water Cherenkov (HAWC) γ-ray observatory, we perform an archival search over the past decade of data that HAWC has collected to study the potential coincidences that may appear, both temporal and spatial.

## 2. Detectors

### 2.1 HAWC

HAWC is a ground-based gamma-ray detector located on the slopes of the Sierra Negra volcano in Mexico, at an altitude of 4,100 meters [6]. Designed to observe extensive air showers produced by high-energy gamma rays and cosmic rays interacting with the Earth's atmosphere, HAWC consists of an array of over 300 large water Cherenkov detectors. Each detector contains purified water and photomultiplier tubes (PMTs) that capture the Cherenkov light emitted by secondary charged particles traveling faster than the speed of light in water. With a wide field of view ( 2 steradians) and a near-continuous duty cycle, HAWC is uniquely suited for monitoring transient and extended sources of gamma rays in the TeV energy range, playing a crucial role in time-domain and multi-messenger astrophysics.

### 2.2 IceCube

The IceCube Neutrino Observatory is a cubic-kilometer-scale detector embedded deep within the Antarctic ice at the South Pole, designed to detect high-energy neutrinos from astrophysical sources. It consists of over 5,000 digital optical modules (DOMs) deployed along 86 vertical strings, spanning depths from 1,450 to 2,450 meters below the ice surface [7]. These modules detect the faint Cherenkov light emitted when neutrinos interact with the surrounding ice, producing secondary charged particles that travel faster than the local speed of light. IceCube is sensitive to neutrinos across a wide energy range, from tens of GeV to several PeV, making it invaluable to multi-messenger astronomy.

## 3. Data

The IceCube data considered from this analysis comes their first alert catalog IceCat-1 [2] along with later alerts that are continuous being posted to the Astrophysical Multi-messenger Observatory Network [1]. These alerts contain the time, quality, location and uncertainty, energy, and other parameters. As discussed in [2], there are 2 types of alerts: bronze and gold. These correspond to a 30% or 50% probability that the neutrino is astrophysical rather than atmospheric. Combining all available alerts, there are 361 alerts that fall within HAWC's field of view (-20 to 60 degrees of declination), and they span from 2011 to 2025. Depending on the event, these alerts can have location uncertainties ranging from tenths (primarily gold) to ones to tens (primarily bronze) of degrees and are scattered roughly isotopically across the sky. For this analysis, we consider only the location of these alerts, while other parameters will be added in the future.

---
[1] https://gcn.gsfc.nasa.gov/amon_icecube_gold_bronze_events.html

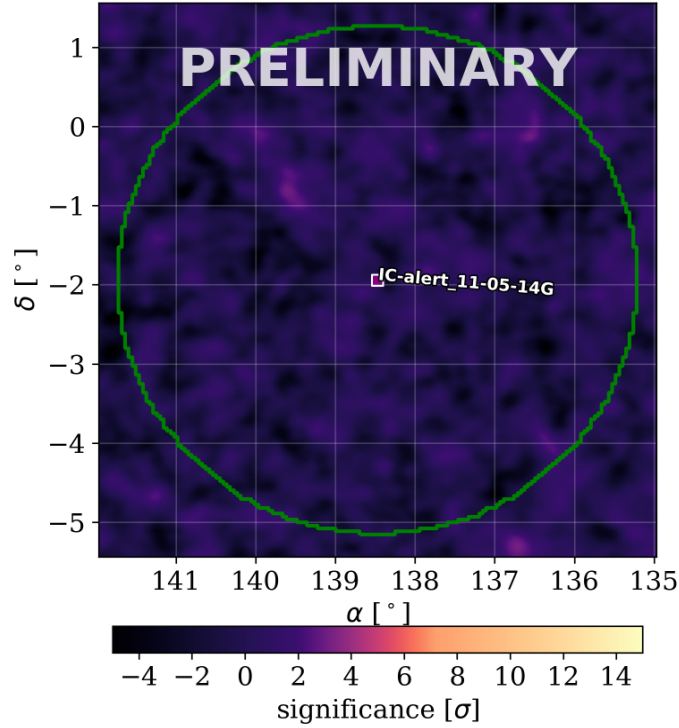

**Figure 1:** An example of an IceCube alert and its uncertainty region. The significance map used here is the full HAWC dataset and indicates that no steady state sources are present within the alert's ROI.

HAWC's data consists of approximately 3000 days of data. This data can either be accessed as a complete set (see [9]) or in daily maps. An example of an IceCube alert overlaid with the full HAWC dataset is given in Figure 1. For this analysis, both data sets are used, with the full dataset giving the IceCube alert's proximity to a steady-state $\gamma$-ray source, and the daily maps indicating potential $\gamma$-ray flares occurring either spatially or temporally coincident with the alert's Region of Interest (ROI).

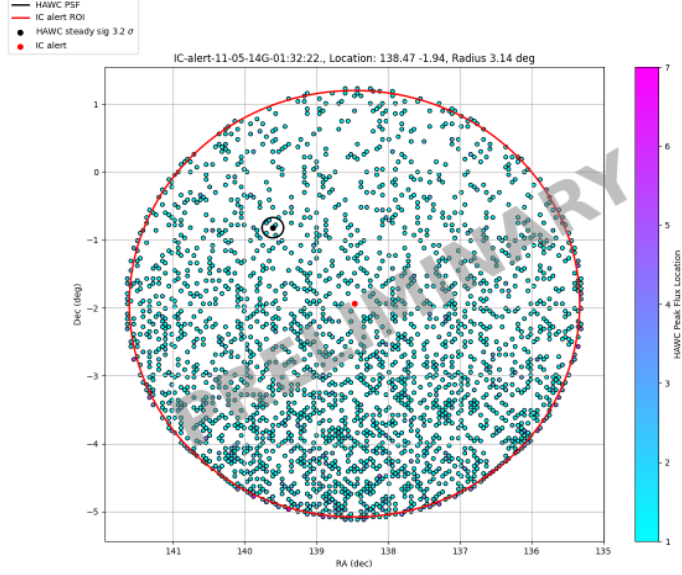

**Figure 2:** A map of the locations for each day's max flux values location. The red dot and circle indicate the IceCube alert's location and uncertainty, while the black point indicates HAWC's max full dataset significance inside the alert's ROI. In this case, no detection is found

## 4. Methodology

To analyze HAWC's daily maps, the following procedure is used. First, daily significance maps are made for every day of available data between November 26, 2014 and Jan 15, 2024 (current end of the available daily maps). These maps are made with a power law spectral assumption, seen in Equation 1.

$$\frac{dN}{dE} = N_o \left( \frac{E}{E_p} \right)^{-\gamma} \tag{1}$$

For significance map generation, the index $\gamma$ and pivot energy $E_p$ are both fixed at 3.0 and 7 TeV respectively while the flux normalization $N_o$ is fitted based on the reconstructed energy $E$ [9]. The index assumption of $\gamma = 3.0$ will be discussed in Section 5, but $\gamma = 2.0$ maps are being tested currently and preliminary results for those will be presented during our talk. For every daily map, full sky (RA = [0, 360], DEC = [-20, 60] deg) are made.

Once the maps are made, the individual IceCube alerts are considered. For each alert, the location (in RA, DEC) and corresponding 90% containment uncertainties are taken and made into separate ROI's. Then, for each daily map, HAWC's maximum reconstructed flux, uncertainty, and location parameters are taken and saved. The final set of locations can be seen in Figure 2, where the color bar indicates repeated locations in the data set. For sources, both steady-state and recurrent transient, clusters would emerge around them. Examples would include the Crab Nebula and the active galactic nuclei Markarian 421. This method will miss less clear sources that may only appear for a few days. To find these potential sources, the flux values are considered.

### 4.1 Bayesian Block Algorithm

To fit the flux values, a Bayesian Block algorithm (BBA) is implemented. The BBA used is from [1] while the code implementation is slightly adapted from [3]. An in-depth discussion on the BBA is in [1] while a brief summary is presented here. The purpose of a BBA is to find statistically significant deviations in data. It does this by test-fitting a series of blocks to the data set, calculating a loss function with each block fit. Fits with more blocks receive penalty factors to prevent overfitting. It starts with one, then adds a new block, and then other different possible configurations to create a loss function sample space. It then finds the configuration that best minimizes the loss function, based off some sensitivity parameter. This parameter is called the ncp_prior, and it determines how likely the BBA is to prefer one model over another.

### 4.2 Calibrating the BBA

When running the code version of the BBA from [3], the ncp_prior can be passed through as a hyperparameter and calibrated. The goal of this analysis is to achieve a false positive rate (FPR) of 5%. This means that, for 5% of the IceCube alerts, we expect to get more than one block (background only) fitted to our data. To achieve this result, the ncp_prior must be calibrated. One note is that the ncp_prior is currently calibrated off of real data, but simulation studies are being actively pursued.

First, the field of view considered is constricted to be between -20 and 60 degrees. This is divided into 8 bands, each 10 degrees wide. 36 10 degree wide circular ROIs are then defined for each declination bands. These ROIs are centered in the middle of the band, so the first ROI is centered at (5, -15), the next is at (15, -15), and so on. Then, the fitting algorithm described above is used to find the max flux value for every daily map, giving 36 (35 where visible sources are present) sets of 3000 maximum flux values. Concatenating these arrays gives approximately 100,000 data flux values and their uncertainties. This is the data that will be used to calibrate the ncp_prior for each declination band.

The calibration is done using the Monte Carlo method, where the BBA is fit to datasets constructed by randomly sampling from the data collected above. These randomly constructed datasets contain the same number of days as an actual dataset (2917 days). To find the ncp_prior that will give a 5% FPR, ncp_prior from 2.0 to 8.0, stepping by 0.25, are tested. Each ncp_prior is used in fitting 1000 randomly generated lightcurves and, for every non-end point that is found, those points are tallied as false positives. An example of a data lightcurve vs a simulated lightcurve is shown in Figure 3.

The results of this method gives ncp_priors for each declination band. These are given in Figure 4. It is clear that there is some declination dependence though, for the purpose of this analysis, only the peak value at 5.6 is taken.

## 5. Results

With the ncp_prior value found, the full transient analysis is run on all available IceCube alerts. Factoring in the slightly reduced declination range, there are 361 available alerts. Running the analysis yields 71 detections, or a detection rate of 20%. These detections fall into 4 categories: 2

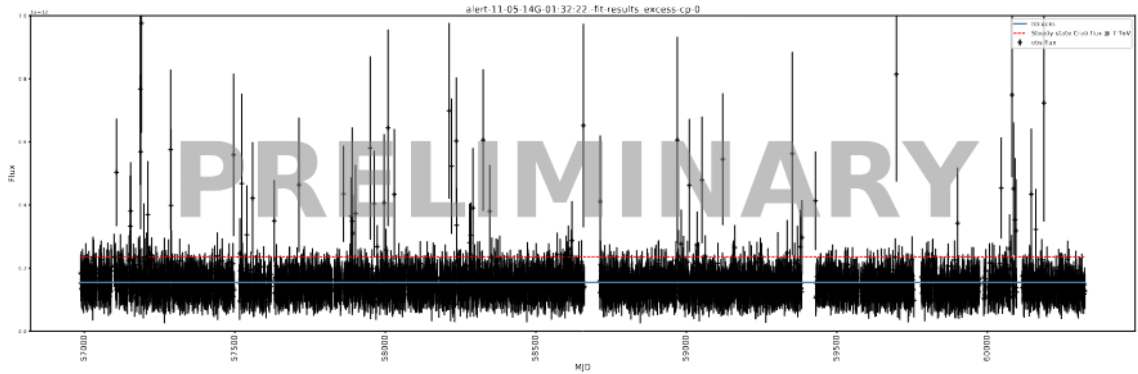

**(a)** An example of a real IceCube alert being fit with the BBA. The black points are the max flux values for alert, with the blue line indicated the BBA fit result. In this case, no detections were found. The red line indicates the steady-state flux of the Crab at 7 TeV, which is the assumption that all the maps used.

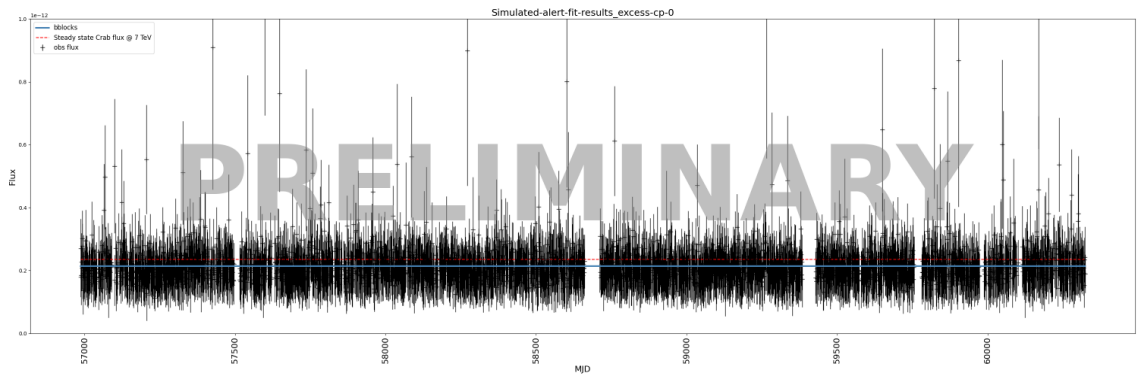

**(b)** A simulated lightcurve from the same declination. Like with the above figure, no detections were found.

**Figure 3:** Two lightcurves with HAWC data.

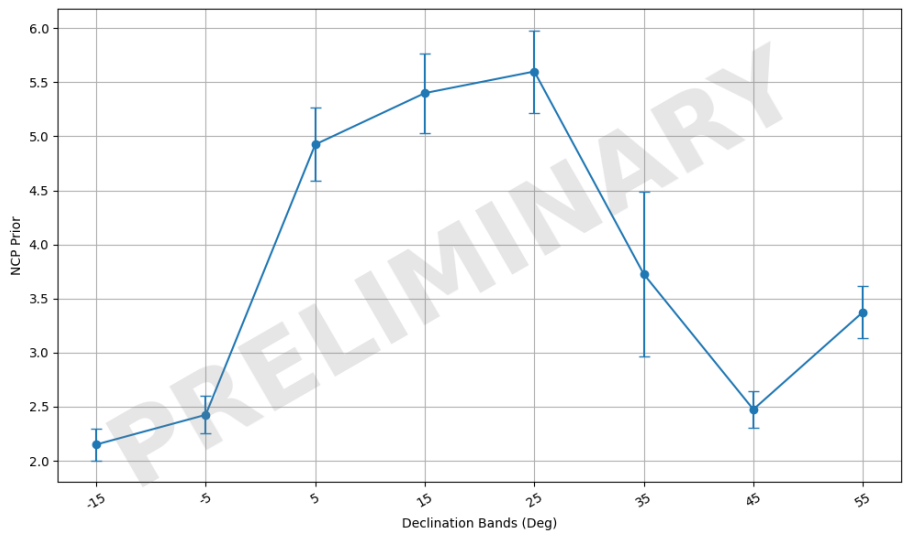

**Figure 4:** ncp_priors that correspond to 5% for each declination. The uncertainties are found from the standard deviation of 20 calibration runs.

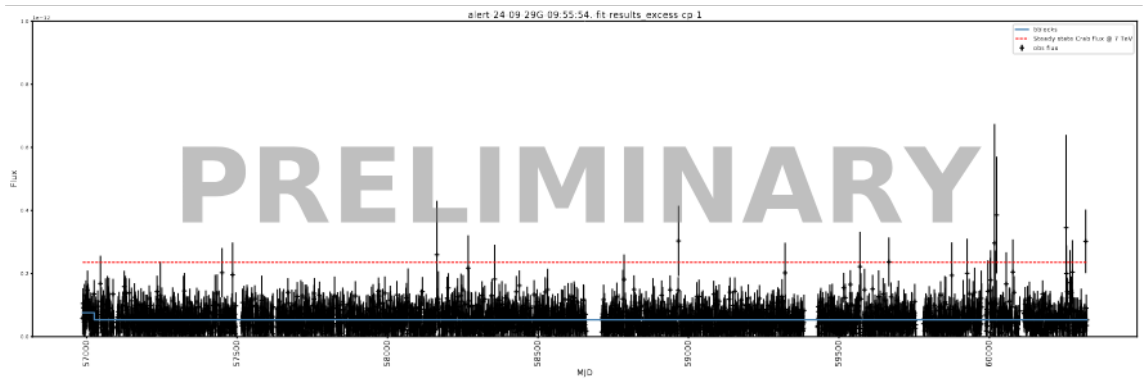

**Figure 5:** An example of a FPR from a high flux region. It is not currently clear where this is coming from.

blocks (1 region of elevated flux) with 50 found, 3 blocks (1 point fit) with 18 of these, 2 that had both of these, and 1 source: Mrk 421. The elevated flux detections all occur within the first year of HAWC's operation and hint at a potential systematic issue with that data. If those are excluded, there are 21 total detections, giving a 5.8% detection rate. This is within expectations. The high flux region is shown in Figure 5.